\documentclass[usegraphicx, useAMS, usenatbib]{mn2e}

\usepackage[T1]{fontenc} 

\usepackage{array}
\usepackage[usenames,dvipsnames,svgnames,table]{xcolor}
\usepackage[thinlines]{easytable}
\usepackage{caption}
\usepackage{subcaption}
\usepackage{amsmath}
\usepackage{hyperref}
\def\oc{\omega_{\rm cdm}}
\def\ob{\omega_{\rm b}}
\def\om{\omega_{\rm m}}
\def\OL{\Omega_{\Lambda}}
\def\taur{\tau_{\rm reio}}
\def\darec{d_A^{\rm rec}}
\def\Al{A_{\rm lp}}
\def\nl{n_{\rm lp}}

\title{Separate Constraints on Early and Late Cosmology}

\author[B. Audren]{B.~Audren$^1$\\
 $^1$ Institut de Th\'eorie des
Ph\'enom\`enes Physiques, \'Ecole Polytechnique F\'ed\'erale de Lausanne, CH-1015, Lausanne, Switzerland}

\begin{document}
\maketitle
\begin{abstract}
Since the public release of Planck data, several attempts
  have been made to explain the observed small tensions with other
  datasets, most of them involving an extension of the $\Lambda$CDM
  Model. We try here an alternative approach to the data analysis,
  based on separating the constraints coming from the different epochs
  in cosmology, in order to assess which part of the Standard Model
  generates the tension with the data. To this end, we perform a
  particular analysis of Planck data probing only the early
  cosmological evolution, until the time of photon decoupling. Then,
  we utilise this result to see if the $\Lambda$CDM model can fit all
  observational constraints probing only the late cosmological
  background evolution, discarding any information concerning the late
  perturbation evolution. We find that all tensions between the
  datasets are removed, suggesting that our standard assumptions on
  the perturbed late-time history, as well as on reionisation, could
  sufficiently bias our parameter extraction and be the source of the
  alleged tensions.
\end{abstract}
\nokeywords

\section{Introduction}

  It is a well established fact that early cosmology history until photon
  recombination is well understood. What happens after this epoch relies
  however on a less solid ground. The nature of dark energy, the details
  of reionisation, the collapse of structures, all this rely on priors
  that are not well tested, or non-linear physics, and thus might bias
  our analysis. 

  As was done originally in~\cite{Vonlanthen2010}, it is possible to
  devise an analysis of the early cosmology parameters that is
  independent of assumptions concerning the late universe, providing
  so-called ``agnostic'' constraints, {\it i.e.} constraints without
  believing in any late-time cosmology model.  In~\cite{Audren2013}, this
  analysis was done with the data sets available at the time, and
  improved in order to be also independent of the CMB lensing
  contamination. It provided a consistency check of our current standard
  model of early-cosmology.

  The purpose of this paper is two-folds. In a first time to update this
  analysis to the current Planck data, while improving the method.
  Then, we propose to use this newly acquired knowledge about the
  early universe and treat it as a measurement of early quantities, in
  the sense that it gives posterior distribution on a set of
  cosmological parameters. From there, one can assume a model for the
  homogeneous late-time evolution, and test the implications of the
  previous measurement on this model, as well as the information
  coming from other probes. The goal would then be to exclusively test
  the merit of the cosmological constant as an explanation for the
  late time acceleration, without any contamination from other
  assumptions. This second point could be extended in the future to
  more general models for late cosmology, involving {\it e.g.}
  neutrino masses or dynamical dark energy.

  The main idea behind this approach is to be able to separate the
  effects of different assumptions on parameter extraction. In order
  to say that the $\Lambda$CDM model is in tension with current
  measurements, one must be sure that this tension is a failure of the
  model to describe the late-time acceleration, and is not due to some
  of our assumptions about structure formation, or about reionisation,
  for instance. We will therefore adopt a simple $\Lambda$CDM model
  for the late homogeneous cosmology.

  With the recent release of Planck temperature anisotropies map, it
  is possible to apply these ideas and see how it affects the
  analysis. Indeed, there are some tensions between the current Planck
  analysis~\cite{PlanckXVI,Planck:2013nga} and the results of other
  cosmological probes: as it has for instance been pointed out
  in~\cite{Verde:2013wza}, current existing constraints on the value
  of $H_0$ disagree with each other. One example is the discrepancy
  between the Planck result: ${H_0 = 67.3\pm
  1.2}$ km~s$^{-1}$~Mpc$^{-1}$, and the Hubble Space telescope
  measurement: ${H_0 = 73.8 \pm 2.4}$ km~s$^{-1}$~Mpc$^{-1}$, at
  $1\sigma$. Although only a $2\sigma$ tension, it might be a sign of
  something wrong in the theoretical assumptions. It has been proposed
  in \cite{Marra:2013rba} that this tension could be partially lifted
  by taking into account the local gravitational potential at the
  position of the observer in the HST measurement, but this effect is
  not enough to sufficiently relieve the tension. Another mismatch
  exists between the value of $\sigma_8$ as probed by the weak lensing
  of the CMB or through the Sunyaev-Zel'dovich cluster count identified
  with Planck - it may advocate for neutrino mass although this is not
  favored by Planck temperature anisotropies spectrum alone.
  A recent proposition states that these anomalies are
  alleviated when analysing in a different way the $217$ GHz
  map~\cite{Spergel:2013rxa}. We are assuming here that the standard,
  publicly available likelihood is correct.
  
  The idea of the paper is to see if one can reduce the observed
  tensions, by assuming only a minimal number of hypotheses. It is
  interesting to check whether one can make all current experiments
  agree with each other, by performing first an ``agnostic'' early
  universe analysis, and then assuming $\Lambda$CDM for the late-time
  homogeneous evolution. At the very least, it would show the
  importance of the missing assumptions, especially in the case of
  studying extended standard models.

  In section~\ref{sec:agnostic}, we will present an improved
  ``agnostic'' analysis method, and discuss its similarities and differences
  with the standard analyses. In section~\ref{sec:late}, we will show
  how to take one further step and derive constraints on a standard
  $\Lambda$CDM model coming from different homogeneous probes. We will
  show and discuss the results in section~\ref{sec:results} and
  conclude in section~\ref{sec:conclusion}.

\section{Agnostic Study}\label{sec:agnostic}

  The main idea on which the so-called ``agnostic''
  study~\cite{Audren2013} relies on is the realisation that, at the
  level of the primary (unlensed) power spectrum, the late-time
  cosmological parameters of some standard scenarios have a clear
  effect. They simply globally rescale in amplitude (through $\taur$
  in the Standard Model), or shift the position of the peaks through a
  rescaling of the multipoles (the action of a cosmological constant
  $\Lambda$, for instance). Both effects behave as mentioned only for
  large multipoles ({\it i.e.} $\ell \geq 50$). By allowing to
  marginalise over these parameters, one should in principle be able
  to extract constraints on early cosmology parameters, independent on
  our assumptions for the late evolution.  This previous analysis
  however had arguably one weakness about the lensing treatment that
  we will address here, after recalling the basic principle of the
  method.

  More mathematically, the effect of the late time parameters on the
  temperature anisotropies power spectrum is that of a global
  rescaling ($C_\ell \to \alpha C_\ell$), and an arbitrary scaling of
  the multipoles ($C_\ell \to C_{\beta\ell}$). Given this freedom, one
  can super-impose exactly two power spectra coming from two universes
  with the same early composition, but different late time
  cosmological parameter (namely the reionisation depth $\taur$ and
  the Hubble parameter $H_0$). This is true as long as one removes
  from the data the information coming from the lowest multipoles,
  $\ell \ll 50$. In practice, when using WMAP data, we tested the
  dependence of the parameter estimation on the starting multipole
  $\ell$ of our analysis, and decided to use $\ell = 100$.

  We do not have this freedom when using Planck likelihood, as the
  starting $\ell$ is fixed. We can, however, restrain from using the
  low multipole likelihood, and exploit only the likelihood for high
  multipole, starting at $\ell = 50$. In the future, we hope to be
  able to specify the starting $\ell$ of the analysis with the next
  releases of Planck likelihood codes.

  As in this previous paper, if one would measure directly the primary
  anisotropies, the set of parameters probed by such an agnostic
  analysis would be the following:

  \begin{align}
    \{\ob, \oc, e^{-2\taur}A_s, n_s, \darec\}
    \label{}
  \end{align}

  Indeed, only the product $e^{-2\taur}A_s$ and the angular diameter distance
  $\darec$ are constrained by the information contained in the CMB
  alone, when we allow for an arbitrary normalisation, and rescaling of
  multipoles.

  Note that this discussion is still valid for other cosmological models where
  only the late evolution is different ({\it e.g.} with spatial curvature,
  non-zero neutrino mass, dynamical Dark Energy, but not for models with
  $N_{\rm eff}$, varying constants, or Lorentz-violating dark energy).

  Since one observes in reality lensed anisotropies, it is crucial to
  treat the effect of lensing on the CMB photons in the same agnostic
  way. In a normal analysis, the lensing potential is generated by the
  same initial power spectrum amplitude and tilt than the one generating
  the perturbations. However, this assumes that $\Lambda$CDM is valid
  for the late-homogeneous evolution, and this assumption should not be
  used here.

  If one wants to be as general as possible, this power spectrum can
  have any shape and amplitude, and should not be the same as the one
  generated by $A_s$ and $n_s$. In~\cite{Audren2013}, two parameters
  $A_{\rm lp}$ and $n_{\rm lp}$ were introduced, for the lensing
  potential, that simply modified the shape of the original spectrum.
  Hence, $A_{\rm lp}=1$ and $n_{\rm lp} = 0$ correspond to the
  standard amount of lensing generated by the underlying power
  spectrum in a $\Lambda$CDM universe. In this parametrisation, the
  meaning of these value changes from one point in parameter
  space to the other, because they are defined with respect to the
  initial power spectrum.

  Instead, in this paper, we reformulate the approach, and we use
  these two parameters $\Al$ and $\nl$ to define, on their own,
  respectively the amplitude and the tilt of the lensing potential,
  and allow to marginalise over them.  In this way, if they are equal
  to $A_s$ and $n_s$, respectively, it will mean that the lensing is
  caused by a late-time $\Lambda$CDM universe. By allowing them to
  vary freely, we do not impose this prior knowledge. We moreover set
  the pivot scale of this lensing potential to coincide with the
  maximum of the lensing potential, which is roughly $k=0.012$1/Mpc
  for Planck data. 

  Finally, we set the effective number of relativistic species $N_{\rm
  eff}$ to $2.03351$, and we take one massive neutrino of a total mass
  of $0.06$ eV, as specified in the base analysis of the Planck study.
  These values impact the prediction of $H_0$ by $0.6$ km s$^{-1}$
  Mpc$^{-1}$ - a significant change considering Planck error bars on
  the Hubble rate.

  We run the Markov Chain Monte Carlo code Monte Python on Planck
  data, in which we only take the high-$\ell$ likelihood (starting at
  $\ell = 50$). We discard the information coming from the WMAP
  polarisation data, and from the low-$\ell$ likelihood. We also
  discard the lensing reconstruction likelihood to avoid making
  hypotheses on structure formation.

  The set of varied parameters is $$\{\ob, \oc, e^{-2\taur}A_s, n_s,
  \darec, \Al, \nl\}.$$
  
  It has to be noted that a similar approach was performed in the
  standard analysis~\cite{PlanckXVI}, with only the lensing amplitude
  being varied (parameter $A_{\rm l}$ in this paper), and not the
  lensing tilt. This was simply done to highlight the fact that the
  CMB alone preferred a value slightly higher than $1$ for this
  parameter. The test was done both with $A_{\rm l}$ defined with
  respect to the initial power spectrum, and with $A_{\rm l}$ defined
  on its
  own\footnote{\url{http://www.sciops.esa.int/wikiSI/planckpla/index.php?title=File:Grid_limit68.pdf&instance=Planck_Public_PLA}}.
  These results were however not further investigated, especially the
  latter.
  
  \begin{figure*}
    \centering
    \hspace{-3.4cm}
    \includegraphics[scale=0.46]{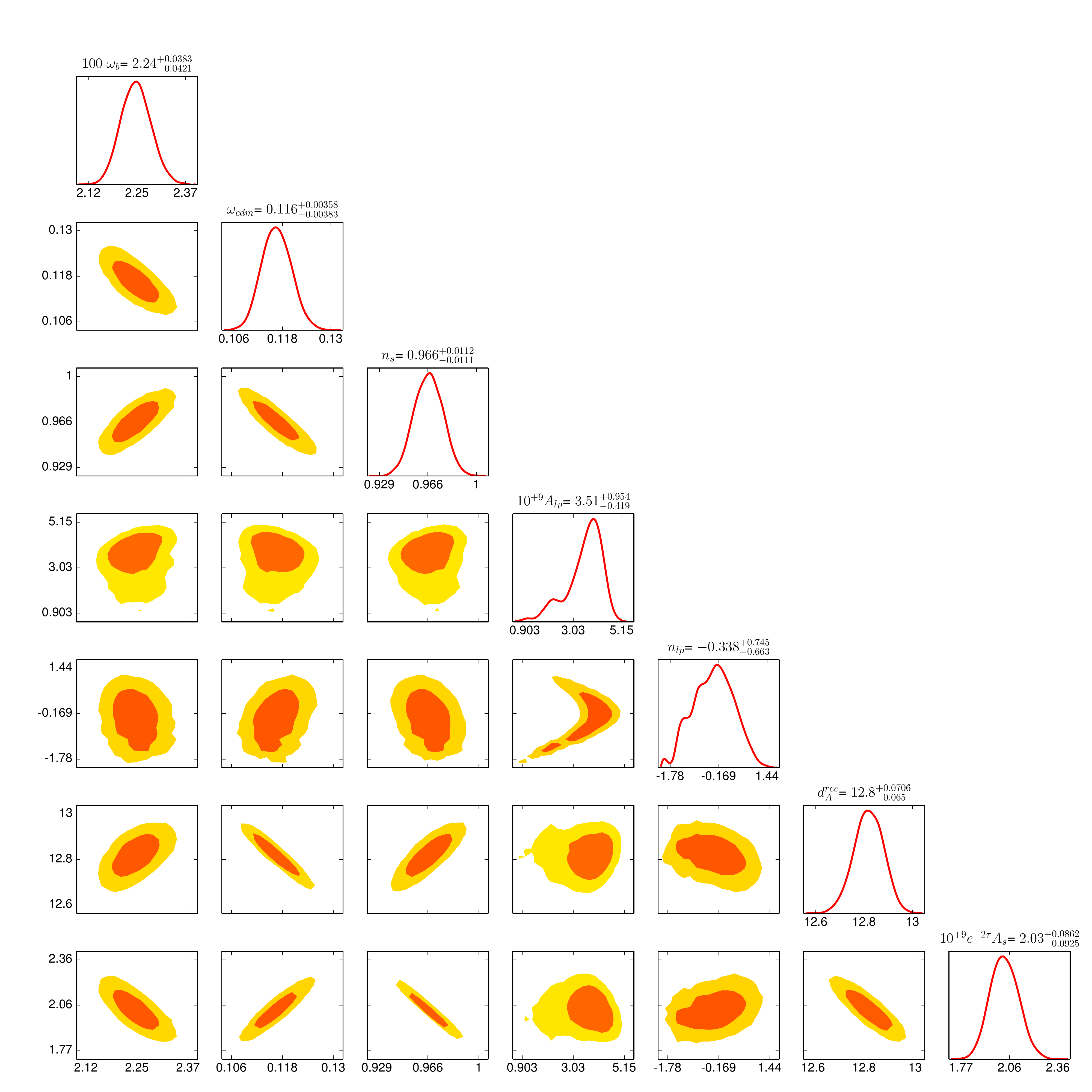}
    \caption{2d posterior distribution for the marginalised
    cosmological parameter of the ``agnostic'' run. The nuisance
    parameters associated with the Planck likelihoods have been
    excluded from the triangle plot for readability reasons.}
    \label{fig:triangle_early}
  \end{figure*}

\section{Constraining the late time homogeneous
  evolution}\label{sec:late}

  We have seen in the previous section how to obtain in principle a
  constraint on early cosmological parameters, with their posterior
  distribution and correlations, in a model-independent way. We want
  now to push the analysis further, and utilise this knowledge to
  determine whether or not $\Lambda$CDM is a good model to explain the
  homogeneous evolution of the late-time universe. 

  The idea is to choose a model for the late-time evolution, namely
  the cosmological constant, and test its merit to explain the
  accelerated expansion. By basing our analysis on the ``agnostic''
  study, we have indeed the possibility to test this single
  assumption, without involving any other one. This approach thus
  differs from the standard one by the fact that we test separately
  the hypotheses of the standard model, instead of evaluating the
  general merit of all of them considered at the same time.

  To this end, we will restrict ourselves to a flat universe, and
  assume further that $\omega_b$ is fixed to its best-fit value from
  the ``agnostic'' study. Indeed, its mean value and its error bar
  are respectively five and ten times lower than the ones of
  $\omega_{cdm}$. Additionally, as $n_s$ will not play any role in the
  homogeneous late-time evolution, it will be considered fixed in this
  study.

  All this results in having the following set of varying parameters

  \begin{align}
    \{h, \Omega_\Lambda\}
    \label{}
  \end{align}

  From these two parameters, and the constant value of $\ob$, we can
  deduce the value of $\oc$. We then test this cosmological set of
  parameters against the following existing data on homogeneous
  cosmology:

  \begin{enumerate}
    \item {\it Direct Measurement}: In \cite{Riess:2011yx}, the authors provide an updated
      measurement of the local value of $H_0$ coming from direct
      distance measurements. It is a simple Gaussian distribution
      centered on $73.8$~km~s$^{-1}$~Mpc$^{-1}$, and a standard
      deviation of $2.4$~km~s$^{-1}$~Mpc$^{-1}$.


    \item {\it Supernovae}: Type-IA supernovae act as standard candles - the luminosity
      produced by their explosion is thought to be a constant,
      regardless of their setup before the explosion. The experiment
      measures then the apparent luminosity of the supernovae, as well
      as its spectroscopic redshift. One can then ask the angular
      diameter distance from the cosmological code, compute the
      luminosity distance with the relation $d_L = (1+z)^2 d_A$, and
      compare it with the observed one. Recalling that

\begin{align} d_L(z_i) = (1+z_i) \int_0^{z_i}
  \frac{dz}{H(z)} = (1+z_i) \int_0^{z_i} \frac{dz}{H_0\sqrt{\Omega_m(1+z)^3 +
  \Omega_\Lambda}}, \label{} \end{align}

      one would expect this probe to be sensitive both to the values of $h$ and
      $\Omega_\Lambda$. However, the likelihood formula uses a simple
      $\chi^2$ formula for each data point, with non zero correlations
      between them, as well as a marginalized nuisance parameter
      accounting for the absolute magnitude of the measurement.
      This leads that only the information on $\Omega_\Lambda$ is
      extracted from this experiment. We used the data
      from~\cite{Amanullah:2010vv} (Union2 data) in this study.

    \item {\it BAO}: The observed Baryonic Acoustic Oscillation scale
      at a given redshift is given by the following ratio:

      \begin{align}
        r_s^{\rm BAO} = \frac{r_s^{\rm drag}}{\left( (d_A)^{2/3}
        (d_R)^{1/3}\right)} \label{} 
      \end{align}

      where $r_s^{\rm drag}$ is the baryon drag scale (Photon and
      baryon are usually considered to decouple at the same time, but
      since there are much less baryons, they actually decouple
      slightly later than the photons (around $z=1000$). This period
      where they are still in equilibrium with the remaining photons
      is called the drag epoch, and the drag scale thus marks the end
      of this epoch). The baryon drag scale is determined by the
      agnostic analysis, with an accuracy better than $1\%$.

      $d_A(z_{\rm survey})$ and $d_R(z_{\rm survey}) = z_{\rm survey}/h(z_{\rm
      survey})$ are respectively the angular diameter and radial
      distances, measured at the redshfit of each galaxy.  The
      denominator consists of the geometric mean of these two
      quantities, that both vary strongly with $h$ and
      $\Omega_\Lambda$. The BAO likelihood is then simply build as a
      $\chi^2$ formula on every measured point. The data used comes
      from 6dFGRS~\cite{Beutler:2011hx},
      SDSS-II~\cite{Padmanabhan:2012hf} (Data Release 7),and
      BOSS~\cite{Anderson:2012sa} (Data Release DR9).

      The standard analysis of the BAO data relies on computing the
      baryon drag scale, as well as angular and radial distances at a
      given redshift, for each point in the parameter space during the
      parameter extraction. We adapted this method to consider the
      baryon drag scale as a measured quantity, coming from the
      agnostic study, with a best-fit and an error. We add both the
      measurement error from the BAO data and this measurement error
      from the agnostic study in quadrature, and keep the simple
      $\chi^2$ formula. It has to be noted that the baryon drag scale
      is measured with a precision of $0.5\%$, so the error is
      dominated by the BAO error.

    \item {\it Time Delay}: Quasar Time-Delay measurements probe cosmological parameters
      through the time delay between different images of
      gravitationally strongly lensed quasars. The chosen quasars have
      a highly intrinsic variable light curve, which is then observed
      coming from separate positions, and thus having travelled
      through different path. The time-delay between the different
      images accounts for differences in the path, but also from the
      different Shapiro delays induced by the lensing galaxy (for an
      in-depth explanation of the measurement, see
      \cite{Tewes:2012gs}). This time-delay distance $D_{\Delta t}$ is
      defined as follows:

      \begin{align}
      D_{\Delta t} = (1+z_d)\frac{D_dD_s}{D_{ds}},
        \label{}
      \end{align}
      where $D_d$ is the angular diameter distance to the lens, $z_d$
      the redshift of the lens, $D_s$ the angular diameter distance to
      the source and $D_{ds}$ the angular diameter distance between
      source and lens. The data we used for this study is taken from~\cite{Suyu:2009by}
      and~\cite{Suyu:2012aa}, with a shifted log normal distribution.

  \end{enumerate}

  Finally, we have to take into account the information coming from
  the early parameter analysis. To do this, we can realise that this
  first analysis gives the posterior distribution for $\ob$, $\oc$ and
  $\darec$. As seen previously, $\darec$ is a function of $\{h,
  \OL\}$, and $\ob + \oc = \om = \frac{1-\OL}{h^2}$, for a flat
  universe. As mentioned previously, since $\ob$ is measured to a much
  greater precision than $\oc$, and since only the sum of the two is
  involved in the late-time evolution, we fixed $\ob$ to its best-fit
  value. 
  
  We thus use the 2-dimensional posterior distribution of $\darec$ and
  $\oc$ found with an ``agnostic'' analysis to define a multi-Gaussian
  likelihood, which seems a reasonable choice considering
  Fig~\ref{fig:triangle_early}. As these two parameters are found to
  be correlated in this first study, it is crucial that our likelihood
  takes into account this degeneracy. If one did not take into account
  this correlation, one would have lost a factor of two in the
  marginalised error bars. In the next section, this last experiment
  will be referred to as {\it this work}. It corresponds then to the
  constraint coming from Planck data alone, which will be compared
  with the aforementioned probes of homogeneous cosmology.

\section{Results}\label{sec:results}

\subsection{``Agnostic'' early universe results}

  After running MCMC chains with {\sc monte
  python}\footnote{\url{http://montepython.net}}, with a modified
  Metropolis-Hastings algorithm~\cite{Lewis:2013hha}, we obtain the
  best-fit, mean and one-sigma constraints for our ``agnostic''
  parameters, as shown in table~\ref{tab:agnostic}. In this table, we
  compare with the standard results based on the high-$\ell$,
  low-$\ell$ Planck likelihoods as well as the WMAP polarisation. As
  we can see there, there are only minor shifts in central values for
  most of the parameters. The degradation of the error bar is not so
  significant.

  Note that it was not possible to perform a comparison
  of the posterior distribution of the cosmological parameters between
  the agnostic approach and a standard analysis using only the
  high-$\ell$ likelihood. Indeed, this data set alone leaves
  unconstrained a degeneracy between $A_s$ and $z_{\rm reio}$, leading
  to extremely poor convergence. It is only with the inclusion of the
  low-$\ell$ likelihood and the WMAP polarisation that converge is
  reached.

  \smallskip
  
  \begin{table}
  \centering
  \begin{TAB}{|l|c|c|}{|c|ccccccc|}
Parameters & This work & Planck Standard  \\
$100~\omega_{b }$  & $2.243_{-0.042}^{+0.038}$ & $2.205 \pm
0.028$ \\
$\omega_{cdm }$ & $0.1165_{-0.0038}^{+0.0036}$ & $0.1199 \pm
0.0027$ \\
$n_{s }$ & $0.966_{-0.011}^{+0.011}$ & $0.9603 \pm 0.0073$ \\
${d_{A}^{rec} }$ & $12.8_{-0.065}^{+0.071}$ & / \\
$10^{+9}e^{-2 \tau} A_{s }$ & $2.032_{-0.093}^{+0.086}$ & / \\
$10^{+9}A_{{lp} }$ & $3.513_{-0.42}^{+0.95}$ & / \\
$n_{{lp} }$ & $-0.338_{-0.66}^{+0.74}$ & / \\
 \end{TAB}
 \caption{The ``agnostic'' analysis gives a value for $-\ln{\cal
 L}_\mathrm{min} = 3895.34$, and a minimum $\chi^2=7791$, to compare
 with $\chi^2=7797.91$ from the standard analysis - an expected
 improvement considering the two additional parameters.}
 \label{tab:agnostic}

\end{table}

  \subsection{Late time universe results}

  The most important differences with respect to the standard analysis
  start to appear when analysing the late-time universe with minimal
  assumptions. 

  The mean values for the $H_0$ parameter, for all the different
  experiments, are summarised in table~\ref{tab:late-lcdm}. One can
  notice on the first hand that, in addition to a wider error bar on
  the Hubble parameter, the central value being significantly different
  than the one from the published Planck analysis. The central value
  and marginalised posterior distribution at 1$\sigma$ are $H_0 = 69.8
  \pm 1.9$km s$^{-1}$ Mpc$^{-1}$, to contrast with $H_0 = 67.3\pm
  1.2$km s$^{-1}$ Mpc$^{-1}$. Note first that the two values are in
  agreement at the level of 1$\sigma$, and that the discrepancy in
  both central value and marginalised width can be attributed to {\it
  i)} the contaminating effect coming from all the underlying
  assumptions usually done when considering the standard $\Lambda$CDM
  universe - mainly reionisation and the growth of structures
  affecting the CMB lensing, and {\it ii)} the information coming from
  discarded region in multipole space.

\begin{table*}
  \centering
  \begin{tabular}{|c|ccccc|}
    \hline
    & Direct Measurement & Supernovae & Time Delay & BAO & This Study\\
    \hline
    $H_0$ & $73.8\pm2.5$ & $65 \pm 27$ & $71.5 \pm 2.9$ & $69.8 \pm 3.5$ &
    $69.8 \pm 1.9$ \\
    \hline
  \end{tabular}
  \caption{Mean values and 1$\sigma$ deviations for $H_0$, for the
  five different late time homogeneous cosmology probes, assuming a
$\Lambda$CDM scenario.}
  \label{tab:late-lcdm}
\end{table*}

\begin{figure}
  \centering
  \includegraphics[scale=0.45]{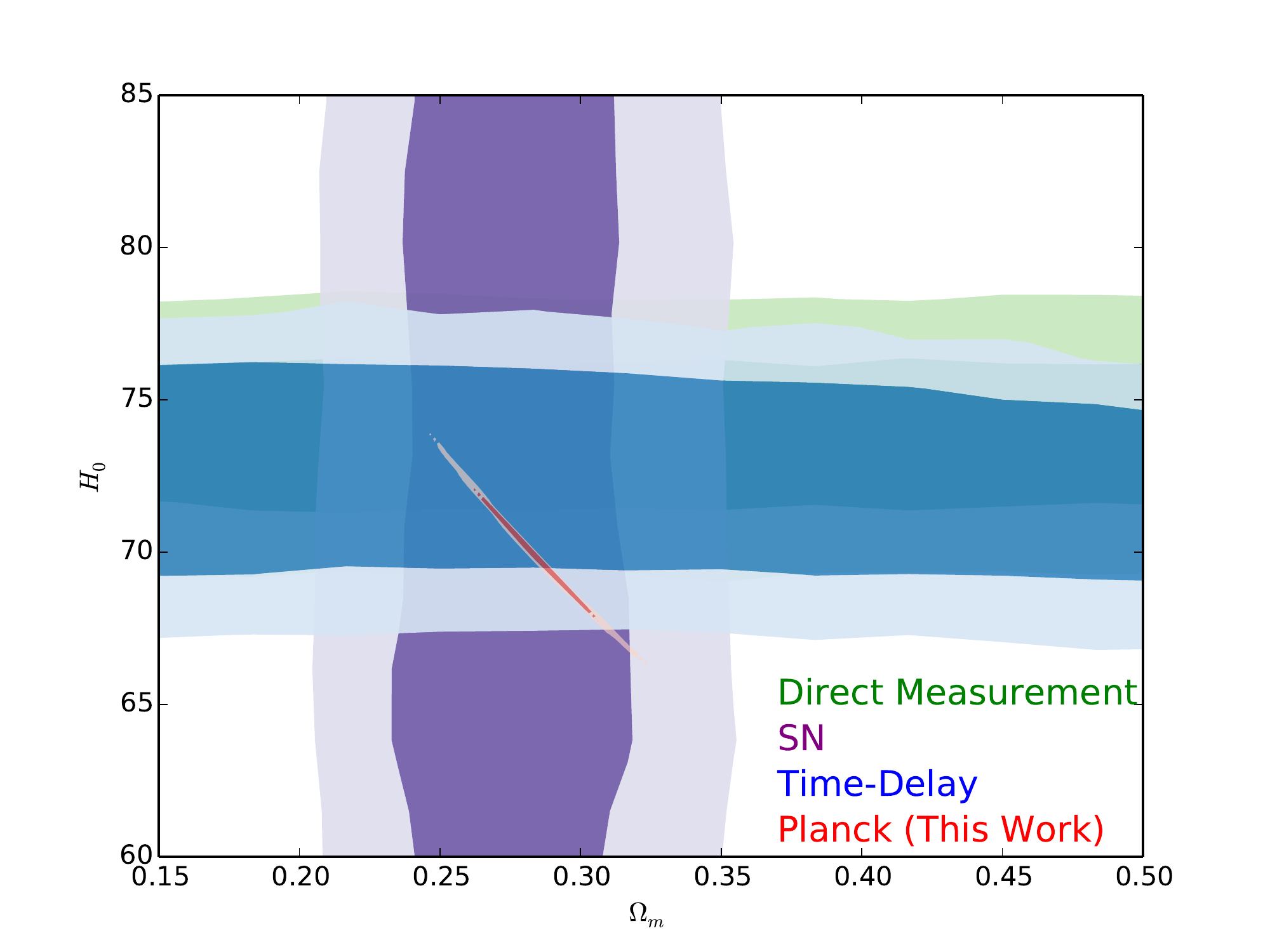}
  \caption{Hubble Space Telescope, Supernovae and Time-Delay posterior
  distribution in the $H_0$, $\Omega_m$ plane, compared to this work
analysis of Planck data.}
  \label{fig:hst-sn-td-ag}
\end{figure}

\begin{figure}
  \centering
  \includegraphics[scale=0.45]{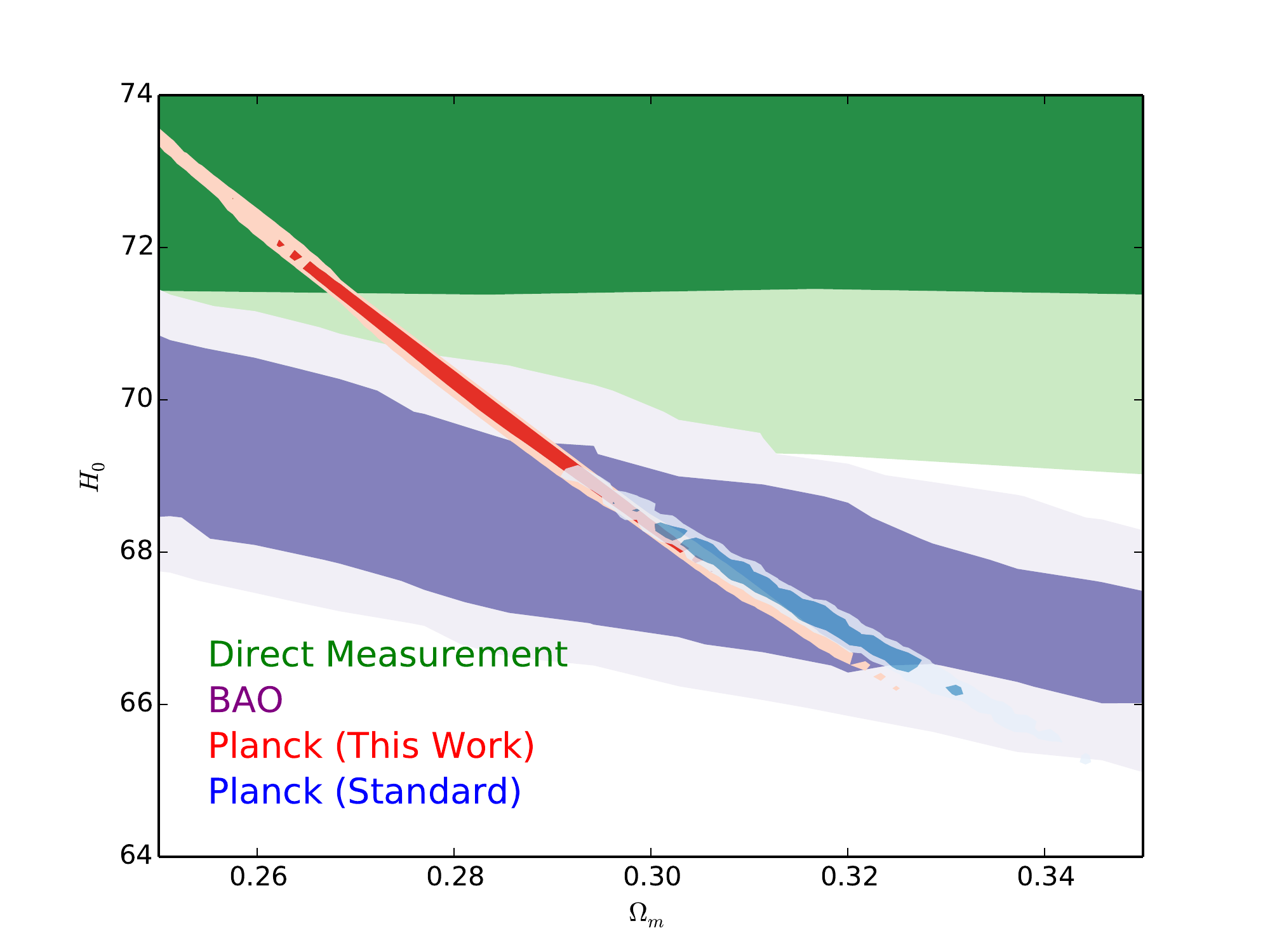}
  \caption{Direct measurement, BAO scale, Planck in this work compared
  to the standard Planck analysis.}
  \label{fig:hst-bao-ag-planck}
\end{figure}

The two dimensional posterior distribution, showing the correlation
between $H_0$ and $\Omega_m$ contains more information, as seen
in figs~\ref{fig:hst-sn-td-ag} and \ref{fig:hst-bao-ag-planck}.

What we can already notice on fig~\ref{fig:hst-sn-td-ag} that the
analysis presented in this study (red contours) agree well with the
constraints coming from direct measurement, supernovae, and time-delay
of quasars. By zooming in, one can see on
fig~\ref{fig:hst-bao-ag-planck} the particular region of interest. We
left out the supernovae and time-delay constraints out for clarity,
but added the BAO one, and the standard Planck analysis (blue
contours).

We can notice there the agreement between Planck standard result and
this work, which has simply a much broader distribution along the
degeneracy between the two parameters. Our more conservative analysis
impacted the sensitivity to the cosmological parameters, thus
reconciling Planck data with the direct measurement of $H_0$, while
maintaining the agreement with the BAO scale measurement.

One can also notice that any tension between the datasets has been
lifted in this proposed analysis. Our claim is that our assumptions on
the perturbed late-time universe, as well as on reionisation, are able
to bias our prediction and could be the source of this observed
tension. 

\section{Conclusion}\label{sec:conclusion}

We devised a way to test the merit of the cosmological constant as the
source of our late time evolution, independently of structure
formation. To this end, we performed a model-independent analysis of
the early universe data coming from the Planck temperature
anisotropies measurement, to obtain so called ``agnostic'' constraints
on the early parameters. We then utilised this knowledge to constrain
the parameter space of $\{H_0, \Omega_m\}$, and compared this analysis
with other experimental probes of the homogeneous late-time universe. 

We showed that, in contrast with the standard analysis, this study
reconciles the local measurement of $H_0$ and Planck data, without
sacrificing the agreement with the other datasets.

Our analysis demonstrated that some of the less often tested
assumptions behind the Standard Model of Cosmology, like the
reionisation history and the growth of structures, can play an
important role in the determination of the posterior distribution of
its parameters. It seems striking that the effects described here are
on par with existing propositions for evidence for new physics in
Planck data.

It is at the best of our knowledge not possible to further refine and
pinpoint which assumption in particular is biasing the most the standard
analysis in the direction of lower $H_0$ values. Indeed, to achieve
this goal, one would need to compare this ``agnostic'' analysis of the
high-$\ell$ likelihood with a standard analysis of the same. However,
as discussed above, the presence of a large degeneracy between $A_s$
and $z_{\rm reio}$ prevents the convergence of the parameter
extraction in this case.

It is not our point to pretend that this method is a better way to
reconcile the datasets than any other proposition, but rather to
highlight the importance of testing as thoroughly as possible every
underlying assumptions of our standard model of cosmology.

This only further highlights the fact that the Planck satellite opened the
doors of a precision era in our field. Such effects were previously
considered unimportant, because of the lack of resolution of past
experiments. With access to such a high sensitivity experiment, a
better understanding of the underlying assumptions behind our Standard
Model, notably the role of reionisation and structure formation, seems
to be in order.

\section*{Acknowledgments}

  We would like to thank Julien Lesgourgues for crucial comments and
  suggestions about the analysis and the draft, as well as Antony
  Lewis for his pertinent remark on the lensing potential amplitude.
  This project is supported by a research grant from the Swiss
  National Science Foundation.

\bibliographystyle{mn2e}
\bibliography{separate}

\end{document}